\documentclass[superscriptaddress,reprint,aps,pra]{revtex4-1}
\usepackage{graphicx}
\usepackage[margin=0.7in]{geometry}
\usepackage{amsmath}
\usepackage{multirow}
\usepackage{float}
\usepackage[usenames,dvipsnames]{xcolor}

\newcommand{\super}[1]{\ensuremath{^{\textrm{#1}}}}
\renewcommand{\vec}[1]{\ensuremath{\textbf{#1}}}
\renewenvironment{acknowledgments}{\textbf{Acknowledgments:}}{}

\setcitestyle{super}

\begin{document}
\title{Electric fields and substrates dramatically accelerate spin relaxation in graphene}

\def\RPIPhy{Department of Physics, Applied Physics and Astronomy, Rensselaer Polytechnic Institute, Troy, NY 12180, USA}
\def\LANL{Theoretical Division, Los Alamos National Laboratory, Los Alamos, NM 87545}
\def\RPIMSE{Department of Materials Science and Engineering, Rensselaer Polytechnic Institute, Troy, NY 12180, USA}
\def\UCSC{Department of Chemistry and Biochemistry, University of California, Santa Cruz, CA 95064, USA}
\def\EqualContrib{These authors contributed equally}

\author{Adela Habib}\thanks{\EqualContrib}\affiliation{\RPIPhy}\affiliation{\LANL}
\author{Junqing Xu}\thanks{\EqualContrib}\affiliation{\UCSC}
\author{Yuan Ping} \email{yuanping@ucsc.edu}\affiliation{\UCSC}
\author{Ravishankar Sundararaman}\email{sundar@rpi.edu}\affiliation{\RPIPhy}\affiliation{\RPIMSE}
\date{\today}

\begin{abstract}
Electrons in graphene are theoretically expected to retain spin states much longer than most materials, making graphene a promising platform for spintronics and quantum information technologies. 
Here, we use first-principles density-matrix (FPDM) dynamics simulations to show that interaction with electric fields and substrates strongly enhance spin relaxation through scattering with phonons.
Consequently, the relaxation time at room temperature reduces from microseconds in free-standing graphene to nanoseconds in graphene on hexagonal boron nitride (hBN) substrate, the order of magnitude typically measured in experiments.
Further, inversion symmetry breaking by hBN introduces a stronger asymmetry in electron and hole spin lifetimes, than predicted by the conventional D'yakonov-Perel' (DP) model for spin relaxation.
Deviations from the conventional DP model are stronger for in-plane spin relaxation, resulting in out-of-plane to in-plane lifetime ratios much greater than 1/2 with a maximum close to the Dirac point.
These FPDM results, independent of symmetry-specific assumptions or material-dependent parameters, also validate recent modifications of the DP model to explain such deviations.
Overall, our results indicate that spin-phonon relaxation in the presence of substrates may be more important in graphene than typically assumed, requiring consideration for graphene-based spin technologies at room temperature.
\end{abstract}

\maketitle
\section{Introduction}
Manipulating spin states of electrons in materials is the emerging frontier in both classical `spintronic' and quantum information technologies.\cite{Recher2010GraphSpin-qubit, wei2014graphSpintronics}
A key requirement for such technologies is that spin states survive during transport over device length scales or for long enough time to carry out a minimum number of quantum operations.
Consequently, an important metric for materials is the spin relaxation time ($\tau_s$), accounting for all relevant scattering mechanisms including intrinsic processes such as spin-phonon relaxation against lattice vibrations that contribute even for an ideally defect-free material, as well as extrinsic processes such as spin-impurity scattering that depends on material quality.

Graphene is an exciting material platform for spintronics and spin qubits because of predictions that spin states could survive microseconds at room temperature if limited only by the intrinsic spin-phonon relaxation mechanism.\cite{geim2007GraphAppPromise, Recher2010GraphSpin-qubit, wei2014graphSpintronics}
However, experimental measurements typically find spin lifetimes at the nanosecond scale, at least two orders of magnitude below the predicted intrinsic limit.\cite{han2011GraphSiO2SpinTemp, zomer2012ChBNlongdistSpinhighmob, han2012SpinRelaxGraphSiO2ChargImpurNoAffect10K, lundberg2013GraphSiO2MagenticDefectExptLowTemp, drogeler2014grhBNSingleFewLayer300K, drogeler2016ChBNSpinValve, raes2016GraphSiO2AnisObliqePrecession, ringer2018GraphSiO2xHanle, kamalakar2015CVDGraphSiO2TempnCarr, kamalakar2020CVDgraphSiO2}
Beyond the overall lifetimes, an additional test of spin relaxation mechanisms is the ratio between lifetimes of spins perpendicular to the plane of graphene ($\tau_{s\perp}$) to those of spins in the plane of graphene ($\tau_{s\parallel}$). 
Most theoretical studies predict the ratio $\tau_{s\perp}/\tau_{s\parallel}$ to be exactly 1/2, which corresponds to a spin-orbit (SO) effective magnetic field that is completely in-plane.\cite{ertler2009graphSubsSpinTheor, vicent2017graphSpinTempCorrugation}
In contrast, measured values for this ratio strongly exceed 1/2 with typical values from 0.7 to 1.1.\cite{tombros2008graphAnistropy, guimar2014ChBNEfieldAnistropy, raes2016GraphSiO2AnisObliqePrecession, ringer2018GraphSiO2xHanle}
The discrepancies in both the overall lifetimes and the ratio between out-of-plane and in-plane spin relaxation underpin an ongoing debate about which spin relaxation mechanism dominates in graphene-based devices.\cite{avsarGraphSpinReview2020}

Previous theoretical studies based on model Hamiltonians and conventional spin relaxation models such as the D'yakonov-Perel' (DP) model have investigated the role of several possible non-idealities including flexural phonons and substrate-induced corrugations, but found these effects to be insufficient to explain the two order-of-magnitude discrepancy in spin lifetimes.\cite{fratini2013graphFlexuralAnis2To3Thoer, ertler2009graphSubsSpinTheor, vicent2017graphSpinTempCorrugation}
Several extrinsic sources have been proposed to play a major role, including electron-hole puddle relaxation dynamics,\cite{tuan2016graphehPuddle} coupled interaction of spin and pseudospin in the presence of adatoms,\cite{tau2014graphPseudoSpin} grain-boundaries in polycrystalline samples,\cite{stephan2019PolyCrystGraph} and resonant magnetic impurities from polymer residues. \cite{kochan2014GraphResonantMagneticTheor}
Recent theoretical work using a modified version of the DP model has also shown that changes to the SO field predicted by density-functional theory (DFT) calculations of graphene on hBN and with applied electric fields can strongly impact the spin relaxation time.\cite{zollner2019GraphhBNAbInitioSpinRelax, zollner2021GrhBNSpinDP}
This modified DP model predicts large anisotropy ratios, e.g. $\sim$4 near the Dirac point, depending on parameters such as inter-valley scattering rates as inputs to the model.
Consequently, \emph{ab initio} predictions of spin relaxation without symmetry-specific assumptions and material-dependent parameters would be invaluable in validating such models and extending predictive capability to new material systems.

Here, we present first-principles calculations of spin-phonon relaxation in the presence of electric fields and substrates in graphene, and identify deviations from or agreements with simplified spin-relaxation models such as the DP model.
We simulate spin relaxation from first-principles density matrix (FPDM) dynamics that include self-consistent spin-orbit coupling and electron-phonon interaction matrix elements from DFT.
We first show that the SO field changes from completely in-plane for free-standing graphene with an applied electric field, to having a prominent out-of-plane component when placed on an hBN substrate, which explains a part of the increase in spin lifetime anisotropy ratio from 1/2.
We then show that the FPDM dynamics simulations with first-principles electron-phonon interactions further increase the anisotropy ratio and find it to be more sensitive to external electric fields, as compared to the conventional DP model.
Finally, we show that recent modifications of the DP model\cite{Cummings2017grapheneTMDsIntervalleyContrib} capture the electron-hole asymmetry and anisotropy of spin relaxation in graphene on hBN in agreement with our parameter-free FPDM simulations.
Altogether, these findings indicate that the intrinsic spin-phonon relaxation mechanism is much stronger in graphene on hBN than typically assumed, and must be overcome to achieve longer-lived spin states in graphene.

\section{Methods}

\subsection{Simulation technique}

To predict spin-phonon relaxation dynamics from first-principles, we employ \textit{ab initio} density-matrix dynamics simulations in a Lindbladian formalism, which we recently showed to accurately predict spin relaxation times in materials with varying electronic structure and symmetry.\cite{xu2020SpinPhononAbinitio,Xu2020AbinitRealTimeSpinGaAs,Xu-Ge}
Briefly, tracing out phonon degrees of freedom from the quantum Liouville equation of the electron-phonon system and applying the Born-Markov approximation\cite{taj2009microscopic} leads to the Lindbladian dynamics,\cite{rosati2015LindbladePh}
\begin{multline}
\frac{\partial\rho_{\alpha_{1}\alpha_{2}}}{\partial t}
= \frac{2\pi}{\hbar N_{q}}
\sum_{q\lambda\pm\alpha'\alpha'_{1}\alpha'_{2}}n_{q\lambda}^{\pm} \\
\times \mathrm{Re} \left[\begin{array}{c}
\left(I-\rho\right)_{\alpha_{1}\alpha'} A^{q\lambda\pm}_{\alpha'\alpha'_{1}} \rho_{\alpha'_{1}\alpha'_{2}} A^{q\lambda\mp}_{\alpha'_{2}\alpha_{2}}\\
-A^{q\lambda\mp}_{\alpha_{1}\alpha'} \left(I-\rho\right)_{\alpha'\alpha'_{1}} A^{q\lambda\pm}_{\alpha'_{1}\alpha'_{2}} \rho_{\alpha'_{2}\alpha_{2}}
\end{array}\right].\label{eq:Lindblad}
\end{multline}
Here, each $\alpha$ denotes an electron wave-vector $k$ and band index $n$ combination, $\pm$ labels absorption and emission of phonons of wave vector $q=\mp\left(k-k'\right)$ and mode index $\lambda$, and $n_{q\lambda}^{\pm}\equiv n_{q\lambda}+0.5\pm0.5$ where $n_{q\lambda}$ is the Bose occupation factor of phonon with frequency $\omega_{q\lambda}$.
Above, $A^{q\lambda\pm}_{\alpha\alpha'} = g^{q\lambda\pm}_{\alpha\alpha'} \delta^{1/2}(\varepsilon_\alpha - \varepsilon_{\alpha'} \pm \hbar \omega_{q\lambda}) \mathrm{exp}(it(\varepsilon_\alpha - \varepsilon_{\alpha'}))$ is the electron-phonon matrix element ($g^{q\lambda\pm}_{\alpha\alpha'}$) along with an energy-conserving $\delta$-function broadened to a Gaussian and time dependence in the interaction picture, where $\varepsilon_\alpha$ are the electron energies.

All electron and phonon energies and matrix elements are calculated on coarse $k$ and $q$ meshes using the JDFTx density-functional theory software,\cite{sundararaman2017jdftx}
and are then interpolated to extremely fine meshes in a basis of maximally localized Wannier functions.\cite{marzari1997maximally,PhononAssisted,GraphiteHotCarriers,NitrideCarriers}
The density matrix dynamics in Eq.~(\ref{eq:Lindblad}) directly describes the time evolution of a thermodynamic ensemble.
The initial ensemble is generated by applying a small perturbation magnetic field that creates an initial spin imbalance.
We then evolve Eq.~(\ref{eq:Lindblad}) in time using an adaptive Runge-Kutta integrator starting from both in-plane and out-of-plane spin-polarized states,  compute the spin expectation values Tr$[\vec{S}\rho(t)]$ for each, and thereby extract the spin relaxation times $\tau_{s\parallel}$ and $\tau_{s\perp}$.
Most importantly, this density matrix formalism naturally describes coherent and incoherent dynamics of an ensemble on the same footing,\cite{xu2020SpinPhononAbinitio} and is therefore not limited only to initial time scales as coherent quantum simulations eg. time-dependent DFT would be.
See Refs.~\citenum{xu2020SpinPhononAbinitio}, \citenum{Xu-Ge}, and \citenum{Xu2020AbinitRealTimeSpinGaAs} for details on the formalism.

\subsection{Computational details}

For the DFT electron and phonon calculations in JDFTx,\cite{sundararaman2017jdftx} we use the Perdew-Burke-Ernzerhof exchange-correlation functional \cite{perdew1996generalized} with fully-relativistic norm-conserving pseudopotentials \cite{van2018pseudodojo} to include spin-orbit coupling self-consistently, and correspondingly use a kinetic energy cutoff of 37 Hartrees (1000~eV) for the wavefunctions.
The lattice constants and internal geometries are fully relaxed using the DFT+D2 correction method for dispersion interactions,\cite{grimme2006vdw} with scale factor $s_6 = 0.5$ appropriate for graphene heterostructures.\cite{GraphiteHotCarriers}
For graphene, this leads to an in-plane lattice constant of 2.465~\AA.
For graphene on hBN, we use a commensurate unit cell and evaluate different stackings of the two layers: AA (B atop C atom and N atop the other C atom) and AB (B atop C atom and N atop hexagonal void of graphene).
We find the AB stacking to have the minimum energy and to be stable with no imaginary phonon modes, with a converged in-plane lattice constant of 2.486~\AA\ and layer separation of 3.28~\AA, all of which agree with previous predictions.\cite{wang2017ChBNReviewStacking, fan2011ChBNStackingLayer}
We use a vacuum separation of 13.22~\AA~and truncated Coulomb potentials\cite{sundararaman2013regularization} to remove periodic interactions in the $z$ direction.

\begin{figure}
\includegraphics[width=\columnwidth]{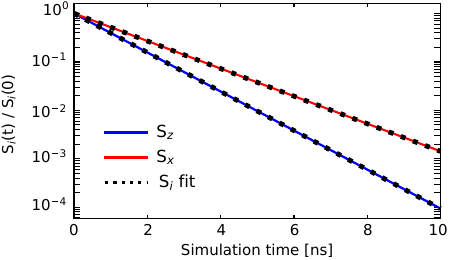}
\caption{Characteristic real-time evolution of in-plane ($S_x$) and out-of-plane ($S_z$) spin expectation values and corresponding fits to extract spin relaxation time, shown here for graphene on hBN with an electron density of 1.2$\times 10^{12}$ cm\super{-2}.}
\label{fig:spinRealTime}
\end{figure}

The DFT calculations are done on `coarse' electronic $\vec{k}$-meshes of $24 \times 24 \times 1$ and phonon $\vec{q}$-meshes of $6 \times 6 \times 1$. 
The density matrix dynamics is performed on a much finer $864 \times 864 \times 1$ mesh for both $\vec{k}$ and $\vec{q}$ using Wannier interpolation of the electron and phonon matrix elements.\cite{giustino2007electron, PhononAssisted, brown2017experimental, NitrideCarriers}
We include the effect of transverse (perpendicular) electric fields $E_\perp = E_z$ as a Stark perturbation Hamiltonian in the Wannier basis, $H_E = e E_z z^{k}_{nn'}$, where $e$ is electronic charge magnitude(1 in atomic units) and $z^{k}_{nn'}$ is the matrix element of the $z$ position operator between electronic bands from the DFT calculation.
The energy-conserving $\delta$-function in Eq.~(\ref{eq:Lindblad}) is implemented as a Gaussian with a finite standard deviation $\sigma$.
We find that the spin relaxation times vary less than $\sim 1$~\% with $\sigma$ below $0.005$~eV, and we set $\sigma = 0.005$~eV for all calculations shown here.

FIG.~\ref{fig:spinRealTime} shows an example of the characteristic time evolution of spin expectation values, starting from an initial spin-imbalanced state created using small magnetic fields that are removed at $t = 0$ as discussed above.
The relaxation times are extracted by fitting $S_i(t) = S_i(0)e^{-t/\tau^{s}_i} \textrm{cos}(\omega t)$, where the $\tau^{s}_{i}$ is the spin relaxation lifetime for Cartesian coordinates, $i=x,y,z$, and $\omega$ is an oscillation frequency related to energy splitting in the band structure. 
For all cases presented here, we find the oscillation periods much longer than relevant spin lifetimes, which is expected because of the extremely small energy splits in graphene, and therefore observe almost perfect exponential decay profiles.
We vary the simulation time based on the system to ensure that the spin has decayed to at least $>50$~\% of its initial value to reliably extract the spin lifetime.
Specifically, depending on the relaxation time, we required simulation times ranging from 500~ps for graphene on hBN to a few $\mu$s ($10^6$~ps) for graphene without fields or substrates.

\section{Results}

\subsection{Internal spin-orbit magnetic field}

\begin{figure}
\includegraphics[width=\columnwidth]{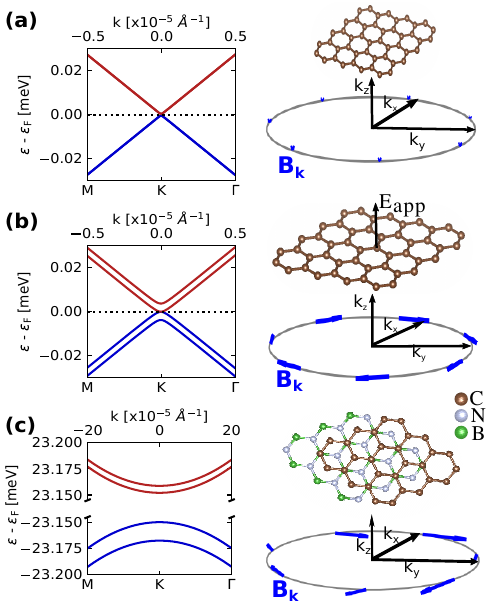}
\caption{Band structure near the K point (left panels) and corresponding $k$-dependent effective SO magnetic field $\vec{B}_{\vec{k}}$ at a Fermi circle 0.1~eV above the conduction band edge (right panels) for (a) free-standing graphene, (b) graphene with electric field, $E_z=0.4$~V/nm, and (c) graphene on hBN. 
The effect of hBN differs qualitatively from that of an electric field, splitting the valence band stronger than the conduction band and introducing an out-of-plane component to $\vec{B}_{\vec{k}}$.
\label{fig:SOF}}
\end{figure}

We begin with an analysis of the spin-orbit (SO) effective magnetic field in first-principles calculations (FIG.~\ref{fig:SOF}), prior to presenting detailed FPDM calculations of spin-phonon relaxation.
Free-standing graphene is inversion symmetric, which implies a zero SO field (FIG.~\ref{fig:SOF}(a)), while electric fields and substrates can break this symmetry and introduce a non-zero SO field.
Conventionally, theoretical studies approximate the impact of the substrate-induced SO field as a Bychkov-Rashba term in the effective Hamiltonian, with a single empirical parameter for the overall coupling strength.\cite{zollner2019GraphhBNAbInitioSpinRelax, vicent2017graphSpinTempCorrugation, fratini2013graphFlexuralAnis2To3Thoer, ertler2009graphSubsSpinTheor}
In the simplest case, the substrate-induced SO field modification is assumed to behave like an electric field applied perpendicular to the graphene plane.\cite{ertler2009graphSubsSpinTheor} This results in a fully in-plane effective SO magnetic field, $\vec{B}_{\vec{k}}$, which is constant in magnitude but varies in direction with wave-vector $\vec{k}$, as seen in (FIG.~\ref{fig:SOF}(b)).
However, realistic substrates can create an atomic-scale electrostatic potential variation on the graphene that is much more complex than a uniform electric field.
Correspondingly, the directionality of the substrate SO field may differ strongly from the simple picture above.

We find that first-principles calculations indeed show qualitative differences in the direction and magnitudes of the SO field between free-standing graphene with an applied electric field and graphene on an hBN substrate (FIG.~\ref{fig:SOF}(c)).
We select hBN as the prototypical substrate for graphene, known to exhibit reduced extrinsic scattering due to low trapped charge densities and a flat profile,\cite{dean12010GraphhBNMobIncrease} and for which several high-quality spin lifetime measurements are available.\cite{guimar2014ChBNEfieldAnistropy, drogeler2016ChBNSpinValve, gurram2018ChBNSpinTransReview}
FIG.~\ref{fig:SOF} specifically shows the internal SO magnetic field $\vec{B}_{\vec{k}}$ for a material with a Fermi level position $\varepsilon_F=0.1$~eV above the conduction band edge.
The SO field is extracted for each point on the Fermi circle from the spin-orbit energy split $\Delta E_{\vec{k}} = g_e\mu_B \vec{B}_{\vec{k}}\cdot\langle\vec{S}\rangle_{\vec{k}}$.
Here, $\langle\vec{S}\rangle_{\vec{k}}$ is the spin expectation value for one of the SO-split bands, and $g_e\mu_B$ is the electron spin gyromagnetic ratio.

Free-standing graphene with no electric field (FIG.~\ref{fig:SOF}(a)) is inversion symmetric and Kramers degenerate, leading to $\vec{B}_{\vec{k}}=0$, exactly as discussed above.
In this case, spin-orbit coupling only introduces a small band gap $\sim 1$~$\mu$eV, as is well-known.\cite{GmitraIntrinsicBandGap, sergej2010GraphGapIntrins}
Applying a transverse electric field to free-standing graphene (FIG.~\ref{fig:SOF}(b)) breaks the inversion symmetry, splits the conduction and valence bands symmetrically and introduces an in-plane azimuthal magnetic field.\cite{ertler2009graphSubsSpinTheor}
We choose a field strength of 0.4~V/nm because it produces a conduction band splitting comparable to that for graphene on hBN (FIG.~\ref{fig:SOF}(c)).
However, note that the valence band splitting is much stronger than that for the conduction band for graphene on hBN, and a band gap of 46~meV opens up.
Most importantly, $\vec{B}_{\vec{k}}$ is no longer in-plane, picking up an out-of-plane component from the substrate interaction. 

\subsection{Spin Relaxation: Carrier Density Dependence}

We next discuss the effect of the changes due to electric field and hBN substrate in the SO field as well as the electron-phonon interaction on the spin relaxation time.
In addition to the direct FPDM simulations detailed above, we also present spin lifetimes calculated using models for two idealized limits: the Elliot-Yafet (EY) mechanism for inversion-symmetric systems,\cite{elliott1954theory, yafet1963g} and D'yakonov-Perel' (DP) mechanism for inversion-symmetry-broken systems.\cite{dyakonov1972spin}
Briefly, in the EY case, SO-based spin mixing facilitates spin-flip transitions between pairs of Kramers degenerate states, leading to a direct correlation between spin-phonon and other electron-phonon relaxation processes, such as $\tau_s \propto \tau_p$, the momentum (carrier) relaxation time $\tau_p$.
In contrast, the DP mechanism involves electron spins precessing between scattering events due to the internal SO magnetic field, resulting in $\tau_s \propto \tau^{-1}_{p}$. The constant of proportionality in these two relations vary in literature because of the simplifications adopted for SO fields.\cite{fabian1998spin, leyland2007oscillatory, vzutic2004spintronics}
Here, for EY, we choose~\cite{Xu-Ge}
\begin{equation}
\left(\tau_{s,i}^{\mathrm{EY}}\right)^{-1} \approx 4\left\langle b_{i}^{2}\right\rangle \left\langle \tau_{p}^{-1}\right\rangle
\label{eq:EY}
\end{equation}
along each Cartesian direction $i=x,y,z$, where $b_{i}^{2} = 0.5 - S_{i}$ is the spin mixing between spin-split states.\cite{vzutic2004spintronics, kurpas2019spin} 
Here, $\left\langle A\right\rangle = \sum_{kn}f'\left(\varepsilon_{kn}\right)A_{kn}/\sum_{kn}f'\left(\varepsilon_{kn}\right)$ is the average of each electronic quantity $A$ near the Fermi surface.\cite{Xu-Ge}
For DP, 
\begin{equation}
\left(\tau_{s,i}^{\mathrm{DP}}\right)^{-1} \approx \left\langle \tau_{p} (\boldsymbol{\Omega}^{2} - \Omega_{i}^{2})\right\rangle,
\label{eq:DP}
\end{equation}
where $\boldsymbol{\Omega} = g_{e}\mu_{B}\vec{B}$ is the Larmor precession frequency of electron spins in the SO field.\cite{vzutic2004spintronics}
We extract the spin mixing ($b_{i}^{2}$) and the internal SO fields ($\vec{B}$) from electronic DFT, and additionally calculate the carrier lifetime ($\tau_p$) from first-principles calculations of electron-phonon scattering.\cite{NitrideCarriers}

\begin{figure}[ht!]
\includegraphics[width=\columnwidth]{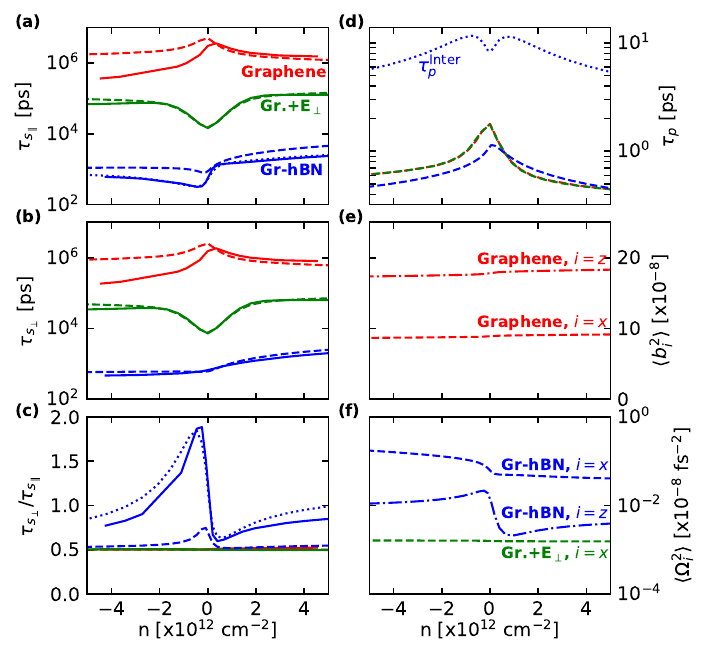}
\caption{(a) In-plane spin relaxation time $\tau_{s\parallel}$, (b) out-of-plane spin relaxation time $\tau_{s\perp}$ and (c) their ratios $\tau_{s\perp}/\tau_{s\parallel}$ in graphene (red), graphene with 0.4~V/nm electric field (green) and graphene on hBN (blue), as a function of carrier density at room temperature using first-principles density matrix (FPDM) dynamics (solid lines) and EY / DP models with first-principles inputs (dashed for conventional DP and dotted for modified DP with inter-valley scattering contribution lines).
(d) Total carrier relaxation times (three dashed lines) and inter-valley only contribution for graphene on hBN (dotted blue line), (e) spin-mixing coefficients and (f) square of spin-orbit fields, all predicted from first-principles for the DP and EY model estimates of spin lifetimes in (a-c).
The EY and DP estimates respectively agree qualitatively for inversion-symmetric graphene and for the remaining inversion-symmetry-broken cases, as expected.
The conventional DP model agrees better for $\tau_{s\perp}$ and understimates the increase of $\tau_{s\perp}/\tau_{s\parallel}$ from 1/2 compared to the FPDM calculations.
Asymmetry between electrons and hole spin lifetimes is also larger in the FPDM predictions (a-c), with the weaker asymmetry in the models primarily from the spin mixing (e) and SO fields (f), while the carrier lifetimes are mostly symmetric (d).
Modifying the DP model to account for inter-valley carrier scattering separately\cite{Cummings2017grapheneTMDsIntervalleyContrib} brings the predictions of $\tau_{s\perp}$ and the anisotropy (blue dotted lines in (a,c)) in much closer agreement with the FPDM results.
\label{fig:tauCarrComp}}
\end{figure}

FIG.~\ref{fig:tauCarrComp} shows our first-principles calculations of spin-phonon relaxation in graphene with and without electric field, and graphene on hBN, each as functions of carrier density, $n$ (positive or negative for electron and hole doping respectively).
The EY model for inversion-symmetric graphene, and the DP model for the cases when inversion symmetry is broken by an electric field or the hBN substrate, agree qualitatively with FPDM predictions, but with some important quantitative differences discussed next.
First, for inversion-symmetric free-standing graphene, the EY model is more accurate for electrons than for holes, for both in-plane and out-of-plane spin relaxation times (FIG.~\ref{fig:tauCarrComp}(a-b)).
The DP model matches FPDM predictions quantitatively for both in-plane and out-of-plane spin relaxation in graphene with electric field, but only for out-of-plane spin relaxation in graphene on hBN.

The discrepancy of the DP model for in-plane spin relaxation in graphene on hBN can be rectified by modifying the DP model.
Briefly, the DP model assumes that the internal magnetic field effectively changes randomly each time the electron scatters.
The in-plane magnetic field $B_\parallel$ rotates over the Fermi circle and covers all in-plane directions, satisfying this condition, in both graphene with electric field and graphene on hBN.
However, the out-of-plane magnetic field $B_z$, which matters only for in-plane spin relaxation and is present only for graphene on hBN, has the same direction within each valley.
Consequently, only inter-valley scattering will change the $B_z$ for a given electron spin.
As proposed in Ref.~\citenum{Cummings2017grapheneTMDsIntervalleyContrib}, this can be captured by changing the DP model from $(\tau_{s,x}^{\mathrm{DP}})^{-1} \approx \langle \tau_{p} (\Omega_y^2 + \Omega_z^2)\rangle$ as given by Eq.~\ref{eq:DP} for in-plane $x$ spins, to
\begin{equation}
\left(\tau_{s,x}^{\mathrm{mDP}}\right)^{-1} \approx \left\langle \tau_{p} \Omega_y^2 + \tau_{p}^{\mathrm{Inter}}\Omega_z^2\right\rangle,
\label{eq:ModifiedDP}
\end{equation}
where $\tau^\mathrm{Inter}_{p}$ is the inter-valley scattering time (dotted line in FIG.~\ref{fig:tauCarrComp}(d)).
This modified DP model agrees with FPDM predictions for in-plane spin relaxation on graphene on hBN (FIG.~\ref{fig:tauCarrComp}(a)).
On the other hand, the EY model discrepancy for holes in graphene requires an analysis of the electron-phonon matrix elements, discussed below in Section~\ref{sec:EphMat}.

The ratio $\tau_{s\perp}/\tau_{s\parallel}$ (FIG.~\ref{fig:tauCarrComp}(c)) is nearly 1/2 for graphene, because the spin mixing is in the same proportion (FIG.~\ref{fig:tauCarrComp}(e)) within the EY mechanism.
This ratio remains unchanged for graphene with an electric field, but now because $\langle\mathbf{\Omega}^2 - \Omega_z^2\rangle = \langle\Omega_x^2 + \Omega_y^2\rangle = 2\langle\Omega_y^2\rangle$, while $\langle\mathbf{\Omega}^2 - \Omega_x^2\rangle = \langle\Omega_y^2\rangle$ since $\Omega_z = 0$ (FIG.~\ref{fig:tauCarrComp}(f)), leading to $\tau_{s,x}^{\mathrm{DP}} = 2\tau_{s,z}^{\mathrm{DP}}$ using Eq.~\ref{eq:DP}.
This ratio deviates substantially from 1/2 only for graphene on hBN (FIG.~\ref{fig:tauCarrComp}(c)) due to the substrate-induced out-of-plane SO field, $\Omega_z \ne 0$.
The conventional DP model (Eq.~\ref{eq:DP}) only captures part of this dramatic effect seen in the FPDM calculations, while the modifications in Eq.~\ref{eq:ModifiedDP} account for the out-of-plane field correctly and agree with the FPDM results in FIG.~\ref{fig:tauCarrComp}(c).

The spin lifetimes decrease with increasing carrier density magnitude in inversion-symmetric graphene (FIG.~\ref{fig:tauCarrComp}(a,b)), but this trend reverses for both inversion-symmetry-broken cases, in agreement with some experiments.\cite{drogeler2016ChBNSpinValve, zomer2012ChBNlongdistSpinhighmob}
The overall spin lifetimes are reduced by one-two orders of magnitude in free-standing graphene by a transverse electric field of 0.4~V/nm, down from microseconds to tens of nanoseconds.
The squared SO field due to an hBN substrate in FIG.~\ref{fig:tauCarrComp}(f) is about 100x larger than the 0.4~V/nm field, further reducing the spin lifetimes in graphene on hBN to the nanosecond scale, comparable to experimental measurements.\cite{drogeler2016ChBNSpinValve, drogeler2014grhBNSingleFewLayer300K, guimar2014ChBNEfieldAnistropy, zomer2012ChBNlongdistSpinhighmob}

Finally, the spin lifetime is mostly symmetric between electrons and holes for graphene with applied electric fields (FIG.~\ref{fig:tauCarrComp}(a, b)).
However, we find hole lifetimes to be typically 2-3x smaller than electrons for both free-standing graphene and graphene on hBN.
On hBN, this asymmetry is captured by the (modified) DP model and stems primarily from the larger spin-splitting and hence SO field in the valence band compared to the conduction band (FIG.~\ref{fig:SOF}(c)), consistent with previous calculations.\cite{zollner2019GraphhBNAbInitioSpinRelax}
Importantly, this effect depends sensitively on the substrate, and even on hBN, could reverse for a different layer stacking.\cite{zollner2019GraphhBNAbInitioSpinRelax}
Consequently, experiments may find electron-hole asymmetries of either sign depending on the substrate and precise structure,\cite{han2011GraphSiO2SpinTemp, avsar2011GraphSiO2Temp, zomer2012ChBNlongdistSpinhighmob, drogeler2016ChBNSpinValve, raes2016GraphSiO2AnisObliqePrecession, kamalakar2015CVDGraphSiO2TempnCarr} and we focus here on the comparison between FPDM predictions and DP model for the specific lowest-energy stacking of graphene on hBN.

\subsection{Role of electron-phonon matrix elements}\label{sec:EphMat}

\begin{figure*}
\includegraphics[width=\textwidth]{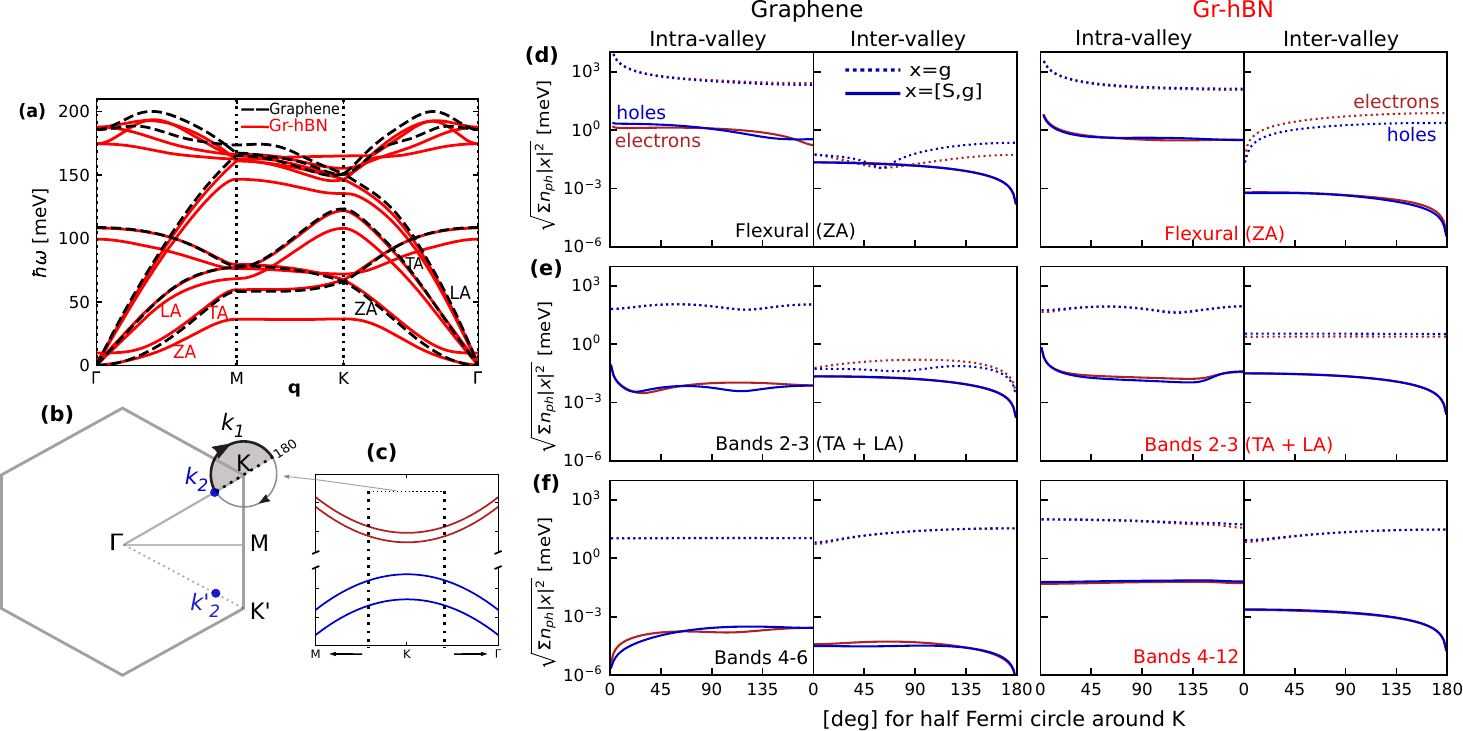} 
\caption{(a) Phonon band structure on high-symmetry $\vec{q}$ path for graphene on hBN (solid red) and graphene (dashed black), labeling three acoustic modes: out-of-plane (ZA i.e. flexural), in-plane transverse (TA), and longitudinal (LA).
(b) Fermi circle at $\varepsilon_F$ = 0.1~eV centered on K used for plotting electron-phonon matrix elements between (c) conduction (electrons, e) and valence (holes, h) band states near K.
(d-f) Electron-phonon matrix elements $|g|$ between $k_1$ half of the Fermi circle (shaded area) shown in (b), and $k_2$ for the intra-valley and $k_2'$ for inter-valley (the row panels for graphene and graphene on hBN). 
The $x$-axis is the angle on the Fermi circle relative to $k_2$, red and blue indicate conduction and valence bands, while solid and dashed lines indicate spin-flip and total e-ph matrix elements, summed over spin-split bands and weighted by phonon occupation factors as detailed in the text. The matrix elements are shown from 0 to 180 degrees that forms the irreducible part of the path by symmetry.
\label{fig:e-phComp}}
\end{figure*}

Above, we were able to explain most features of the FPDM spin relaxation predictions using the EY and (modified) DP models, using first-principles predictions of the spin mixing, internal field and carrier lifetime parameters of these models.
We next analyze the impact of electron-phonon matrix elements on spin relaxation directly, and through these model parameters.
Figure~\ref{fig:e-phComp} compares the phonon dispersion, electron-phonon matrix elements ($x^{q\lambda}_{kn,k'n'} \equiv g^{q\lambda}_{kn,k'n'}$, dotted lines) and spin-flip electron-phonon matrix elements (defined in Ref.~\citenum{xu2020SpinPhononAbinitio}, $x^{q\lambda}_{kn,k'n'} \equiv [S, g^{q\lambda}]_{kn,k'n'}$, solid lines) between graphene on hBN and graphene, separating contributions from the out-of-plane $z$-acoustic (ZA or flexural), in-plane transverse-acoustic (TA) and longitudinal acoustic (LA) phonon modes (FIG.~\ref{fig:e-phComp}(a)).
Each set of matrix elements $x$ above are summed over the spin-split pair of bands for electrons (red) and holes (blue), and weighted by the phonon occupation factors $n_{q\lambda}$ as $\sqrt{(\sum_{nn'\lambda} n_{q\lambda}|x^{q\lambda}_{kn,k'n'}|^2)}$.
To focus on the matrix elements most relevant for spin relaxation in graphene, we select the initial and final $k$ to be on Fermi circles 0.1~eV from the Dirac point (FIG.~\ref{fig:e-phComp}(b,c)).
Then, FIG.~\ref{fig:e-phComp}(d-f) show the matrix elements connecting electronic states at $k_1$ on the circle to $k_2$ and $k_2'$, which are respectively the Fermi-circle points on $\Gamma$-K and $\Gamma$-K' for intra-valley (left sub-panels for both graphene and Gr-hBN columns) and inter-valley (right sub-panels) matrix elements.

First, note common features of the electron-phonon matrix elements in FIGs.~\ref{fig:e-phComp}(d-f) for graphene/hBN and for free-standing graphene.
The spin-flip e-ph matrix elements (solid lines) are orders of magnitude lower than the corresponding total e-ph matrix elements (dotted lines), as expected based on the weak spin-orbit coupling in graphene.
Next, the terms are overall much higher for the ZA phonons (upper panels), followed by the LA + TA phonons (middle panels), and then the remaining modes (lower panels).
Within each panel, the intra-valley terms (right sub-panels for both systems) are much larger than the corresponding inter-valley terms (left sub-panels for both systems), except for the LA + TA phonons.
Both of the above facts are primarily due to the phonon occupation factors, which are largest for the lowest modes (ZA) and for $\vec{q}\to 0$ (intra-valley).
Additionally, the inter-valley spin-flip contributions are exactly zero in graphene at $\theta$ = 180$^{\circ}$ for all phonon modes due to the pseudospin (sublattice) symmetry, while they are still very small for graphene on hBN because the substrate weakly breaks the sublattice symmetry.
Altogether, these points indicate that intra-valley scattering by ZA phonons dominate both the spin-flip and total cases.

Now, consider the spin-flip matrix elements for the EY mechanism of spin relaxation in free-standing graphene, which will be dominated by the intra-valley ZA contributions (left sub-panel in FIG.~\ref{fig:e-phComp}(d) for graphene column).
Averaged over the full Fermi circle, the blue line for holes is $\sim 1.7\times$ higher than the red line for electrons, which will lead to $3\times$ higher spin-flip scattering and hence lower lifetime for holes, as seen in FIG.~\ref{fig:tauCarrComp}(a-b).
However, the total matrix elements for the dominant intra-valley ZA contributions (dotted lines in FIG.~\ref{fig:e-phComp}(d)) are almost equal for electrons and holes, leading to mostly symmetric carrier life times between electrons and holes in FIG.~\ref{fig:tauCarrComp}(d).
This leads to the discrepancy between the symmetric EY estimates and the asymmetric first-principles results for free-standing graphene in FIG.~\ref{fig:tauCarrComp}(a-b).
Essentially, this indicates that the electron-phonon spin-flip matrix elements are not exactly the product of the spin mixing factor and the overall electron-phonon matrix elements, as assumed by the EY model.

Next, the total scattering, rather than spin-flip scattering, contributes to the DP mechanism.
For free-standing graphene, as discussed above, the carrier relaxation rate is dominated by the intra-valley ZA terms and leads to an almost symmetric $\tau_p$ in FIG.~\ref{fig:tauCarrComp}(d), and this does not change appreciably with an electric field.
The situation is similar for graphene on hBN, dominated by intra-valley ZA (dotted lines in FIG.~\ref{fig:e-phComp}(d), right sub-panel in Gr-hBN column), also leading to an almost symmetric $\tau_p$ in FIG.~\ref{fig:tauCarrComp}(d).
The modified DP model (Eq.~\ref{eq:ModifiedDP}) additionally requires the inter-valley carrier scattering time $\tau_p^{\textrm{Inter}}$ for graphene on hBN.
Inter-valley contributions are comparable between ZA and LA+TA (dotted lines in FIG.~\ref{fig:e-phComp}(d-e), left sub-panel in Gr-hBN column), which interestingly exhibit opposite asymmetries between electrons and holes: this cancellation leads to an almost symmetric $\tau_p^{\textrm{Inter}}$ in FIG.~\ref{fig:tauCarrComp}(d).
Altogether, all the relevant $\tau_p$ for DP relaxation in graphene and graphene on hBN turn out to be almost symmetric between electrons and holes, leading to any electron-hole asymmetries being dominated by the internal SO field (FIG.~\ref{fig:tauCarrComp}(f)), as discussed previously.

\subsection{Temperature Dependence}

\begin{figure}
\centering{\includegraphics[width=\columnwidth]{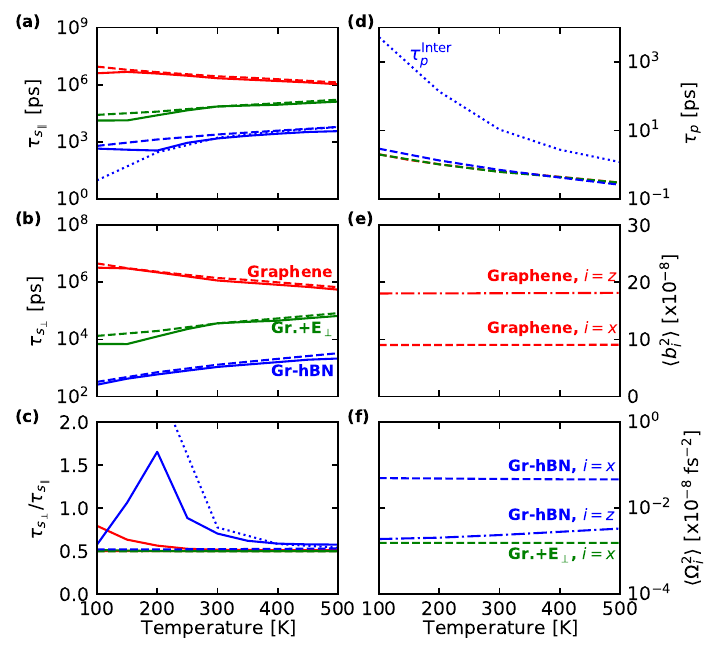}}
\caption{Like FIG.~\ref{fig:tauCarrComp}, but as a function of temperature at fixed electron density, $n = 10^{12}$~cm\super{-2} ($\varepsilon_F\sim 0.1$~eV above Dirac point).
Both in-plane and out-of-plane spin lifetimes (a,b) decrease with increasing temperature for graphene, but increase when inversion symmetry is broken by an electric field or an hBN substrate.
This is expected from the EY model for inversion-symmetric graphene and the DP model for the symmetry-broken cases: the carrier lifetimes and the inter-valley carrier lifetime $\tau_p^\mathrm{inter}$ decrease with increasing temperature in all cases (d), while the spin mixing (e) and internal SO fields (f) are nearly temperature independent.
The EY and conventional DP models agree with the FPDM predictions, except for the in-plane graphene on hBN case, which requires the modified DP model (dotted blue line in (a)) that accounts for inter-valley carrier scattering (dotted blue line in (d)).
\label{fig:spinTempComp}}
\end{figure}

We have so far shown that at room temperature, spin lifetimes exhibit opposite trends with carrier density in inversion-symmetric graphene compared to the case when inversion symmetry is broken by either electric fields or substrates.
The spin lifetimes are also reduced by orders of magnitude, down to the nanosecond scale measured in experiments, in both FPDM predictions and the DP model, primarily due to strong increases of the internal SO field.
Next, we investigate the temperature dependence of spin-phonon relaxation, comparing FPDM predictions to the EY and DP models in FIG.~\ref{fig:spinTempComp}.

Our first-principles density-matrix dynamics calculations show that the spin lifetimes decrease with increasing temperature for free-standing graphene due to increased phonon scattering, both for the in-plane and out-of-plane cases (FIG.~\ref{fig:spinTempComp}(a,b)).
However, when inversion symmetry is broken by either field or substrate, we predict the opposite trend that lifetime increases slightly with increasing temperature.
All these cases follow from the carrier lifetimes $\tau_p$ decreasing with increasing temperature due to increased phonon scattering in all cases (FIG.~\ref{fig:spinTempComp}(d)), with the EY $\tau_s \propto \tau_p$ and the DP $\tau_s \propto \tau_p^{-1}$ (and since spin mixing and internal SO fields (FIG.~\ref{fig:spinTempComp}(e,f)) are nearly temperature independent).

The largest disagreement with the FPDM predictions is for the conventional DP model for in-plane spin relaxation of graphene on hBN, exactly as discussed previously for the carrier density dependence.
Once again, the modified DP model\cite{Cummings2017grapheneTMDsIntervalleyContrib} (Eq.~\ref{eq:ModifiedDP}) fixes this disagreement for temperatures of 200~K and above.
However, below 200~K, the conventional DP model agrees with the FPDM results better than the modified DP model.
This is because the inter-valley scattering time increases sharply with decreasing temperature (dotted line in FIG.~\ref{fig:spinTempComp}d), leading eventually to $\Omega_z \tau_p^\mathrm{Inter} \gg 1$, at which point the DP mechanism no longer operates for the out-of-plane SO field. (DP remains valid for the in-plane SO field with $\Omega_{x,y} \tau_p < 1$.)
In this limit, the conventional DP model that does not account for $\tau_p^\mathrm{Inter}$ then agrees better with the FPDM results.
Due to the above, both DP model versions exhibit monotonic in-plane spin lifetimes and ratios (FIG.~\ref{fig:spinTempComp}(a,c)), while the FPDM results exhibit a maximum ratio around 200~K where $\tau_p^\mathrm{Inter}$ reaches its highest value before scattering becomes too weak for the DP mechanism to take effect.
The ratio remains pinned to 1/2 for free-standing graphene, both with and without an electric field, exactly as discussed previously for the carrier density dependence.

Finally, we emphasize that the predictions shown here are for graphene without defects, accounting only for spin-phonon relaxation.
Previous models accounting for defects and corrugation\cite{ertler2009graphSubsSpinTheor, vicent2017graphSpinTempCorrugation} predict spin lifetimes that weakly decrease with increasing temperature in the inversion-symmetry-broken case as well.
When disorder or impurity scattering dominates, $\tau_p$ will be approximately temperature independent, leading to a nearly temperature-independent $\tau_s$ within the DP model, or even a slightly increasing $\tau_s$ accounting for additional spin-flip scattering by the defects, as seen in some measurements.\cite{avsar2011GraphSiO2Temp, kamalakar2015CVDGraphSiO2TempnCarr, han2011GraphSiO2SpinTemp}
Here, we focused specifically on the strongly temperature-dependent electron-phonon scattering effects in defect-free graphene that required first-principles treatment.

\subsection{Graphene on hBN with applied electric field}

\begin{figure}
\includegraphics[width=\columnwidth]{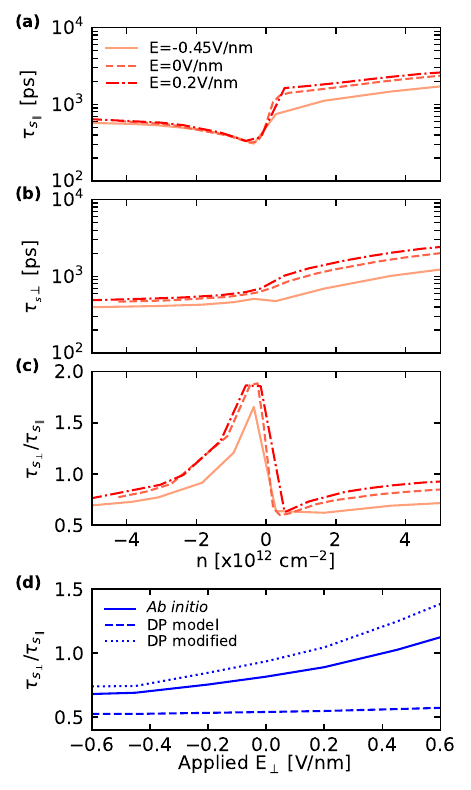}
\caption{First-principles density-matrix dynamics predictions of (a) in-plane spin-phonon relaxtion times, (b) out-of-plane spin relaxation time and (c) their ratio in graphene on hBN as a function of carrier density for various applied electric fields.
(d) Comparison of both DP model estimates to FPDM predictions as a function of electric field at a carrier density of $3.7\times 10^{12}$~cm\super{-2}.
Electric fields may tune the asymmetry between electron and hole spin lifetimes, as well as the ratio between out-of-plane and in-plane spin relaxation.
\label{fig:tauCarrGrhBN}}
\end{figure}

So far, we compared the effect of an hBN substrate on spin-phonon relaxation in graphene to that of an electric field.
Finally, here, we analyze the effect of applying an electric field to graphene on hBN, combining these two effects.
Figure~\ref{fig:tauCarrGrhBN}(a-c) show the spin lifetimes and the out-of-plane to in-plane ratio for graphene on hBN with and without transverse electric fields.
With no field, the spin lifetimes for this lowest energy stacking of graphene on hBN are larger for electrons than holes, as discussed above.
Positive electric fields enhance this asymmetry, while negative electric fields reduce it.

The electric field also modifies the ratio between out-of-plane and in-plane spin lifetimes, increasing it for holes and reducing for electrons as shown in FIG.~\ref{fig:tauCarrGrhBN}(c).
The effect on the ratio is also shown as a function of field at fixed carrier density in FIG.~\ref{fig:tauCarrGrhBN}(d).
Note that the FPDM predictions of the ratio increase strongly with positive electric fields, but the conventional DP model estimates remain close to 1/2 for all electric fields.
As before, the modified DP model accounting for inter-valley scattering fixes this discrepancy and predicts ratios larger than 1/2,\cite{zollner2019GraphhBNAbInitioSpinRelax} in agreement with the FPDM predictions.
The strong deviation of this ratio from 1/2 agrees qualitatively with experimental measurements involving hBN substrates,\cite{guimar2014ChBNEfieldAnistropy} but differ in details because we do not account for defect scattering or the encapsulation of graphene by multi-layer hBN on both sides present in the experiment.

\section{Conclusions}

Using first-principles predictions based on Lindbladian density-matrix dynamics, we have shown that both electric fields and substrates strongly enhance spin-phonon relaxation in graphene by at least two orders of magnitude. 
These calculations provide a reference for spin relaxation, free from mechanistic assumptions specific to the symmetry of the material, such as EY for inversion-symmetric cases and DP for inversion-symmetry-broken cases.
This allows evaluation of both classes of models against FPDM results as inversion symmetry is broken in graphene by either electric fields or substrates.

We find that the conventional models of EY and DP qualitatively agree with the \emph{ab initio} density-matrix dynamics simulations, and show a quantitative deviation that is the largest specifically for in-plane spin relaxtion in graphene on hBN.
This deviation is largely fixed by the modified DP model that accounts for inter-valley scattering in the out-of-plane SO field contributions to in-plane spin relaxation.\cite{Cummings2017grapheneTMDsIntervalleyContrib}
This modified DP model with first-principles inputs also predicts the ratio of out-of-plane to in-plane spin relaxation times to exceed 0.7 for graphene on hBN, in quantitative agreement with the FPDM calculations and qualitative agreement with experiment, in contrast to conventional DP predictions that remain close to 1/2.
However, this modified DP model can not describe the low temperature in-plane spin relaxation of graphene/hBN correctly, due to the inter-valley scattering becoming too weak.

The results presented here suggest that spin-phonon relaxation is more important in graphene on hBN than previously anticipated, especially at room tempreature, with both the overall lifetimes and the ratio in qualitative agreement with experiments.
However, the predictions here remain for an idealized limit accounting only for intrinsic spin-phonon relaxation.
An equivalent first-principles treatment of defect scattering and more realistic substrate geometries, such as gate dielectrics on both sides, will be invaluable to approach quantitative prediction of spin dynamics in graphene.

\begin{acknowledgments}
We acknowledge helpful discussions with Stephan Roche and Aron Cummings.
This work is supported by National Science Foundation under grant No. DMR-1956015.
A.H. acknowledges support from the American Association of University Women (AAUW) fellowship.
Y.P. acknowledges the support from the Air Force Office of Scientific Research under AFOSR Award FA9550-YR-1-XYZQ.
Calculations were carried out at the Center for Computational Innovations at Rensselaer Polytechnic Institute.
\end{acknowledgments}

\bibliographystyle{apsrev4-1}
 \end{document}